# Demonstration of the effect of stirring on nucleation from experiments on the International Space Station using the ISS-EML facility

A. K. Gangopadhyay[1]*, M. E. Sellers[1], G. P. Bracker[2], D. Holland-Moritz[3], D. C. Van Hoesen[1], S. Koch[4], P. K. Galenko[4], A. K. Pauls[2], R. W. Hyers[2], and K. F. Kelton[1]

[1]Department of Physics and the Institute of Materials Science and Engineering, Washington University in St. Louis, St. Louis, Missouri 63130, USA

[2]Department of Mechanical and Industrial Engineering, University of Massachusetts, Amherst, MA 01002, USA

[3]Institut für Materialphysik im Weltraum, Deutsches Zentrum für Luft- und Raumfahrt (DLR), 51170 Köln, Germany

[4]Otto-Schott-Institut für Materialforschung, Friedrich-Schiller-Universität-Jena, 07743 Jena, Germany

*anup@wuphys.wustl.edu

**Abstract**

The effect of fluid flow on crystal nucleation in supercooled liquids is not well understood. The variable density and temperature gradients in the liquid make it difficult to study this under terrestrial gravity conditions. Nucleation experiments were therefore made in a microgravity environment using the Electromagnetic Levitation facility on the International Space Station on a bulk glass-forming $Zr_{57}Cu_{15.4}Ni_{12.6}Al_{10}Nb_5$ (Vit106), as well as $Cu_{50}Zr_{50}$ and the quasicrystal-forming $Ti_{39.5}Zr_{39.5}Ni_{21}$ liquids. The maximum supercooling temperatures for each alloy were measured as a function of controlled stirring by applying various combinations of radio frequency positioner and heater voltages to the water-cooled copper coils. The flow patterns were simulated from the known parameters for the coil and the levitated samples. The maximum nucleation temperatures increased systematically with increased fluid flow in the liquids for Vit106, but stayed nearly unchanged for the other two. These results are consistent with the predictions from the coupled-flux model for nucleation.

# INTRODUCTION

Crystal nucleation[1] and subsequent growth[2] in supercooled liquids (i.e. liquids at a temperature below the equilibrium melting temperature, $T_l$) are the two fundamental processes that determine the solidification microstructure.[3] Studies of these processes, then, occupy a central role in condensed matter physics, materials science, and biology. Both thermodynamic and kinetic factors influence nucleation and growth. In nucleation, spontaneous random processes in the liquid lead to the formation of dense, ordered, regions that are characteristic of the nucleating crystal. The growth of these ordered regions is stochastic, which may grow and shrink. Above a critical size it becomes thermodynamically favored for these aggregates to increase in size, eventually leading to their growth into crystalline solids. Homogeneous nucleation is controlled solely by the intrinsic properties of the liquid and nucleating phases and the amount of supercooling. In contrast, heterogeneous nucleation is catalyzed by foreign objects such as undissolved impurities or the container walls. The minimization of these catalytic sites is then crucial for the studies of homogeneous nucleation; containerless processing, using electrostatic (ESL),[4] electromagnetic (EML),[5] aerodynamic[6] and acoustic[7] levitation, allows such studies. While useful for oxide materials, aerodynamic and acoustic levitations are not recommended for metallic liquids since even the highest purity inert gases often contain impurities that catalyze nucleation.

Nucleation is known to be influenced by pressure,[8] electric[9,10] and magnetic fields,[11] and it has been shown to couple with other phase transitions, such as magnetic transition in a liquid.[12,13] The effect of forced convection on crystal growth has also been studied in a number of recent publications.[14,15] However, the effect of stirring (fluid flow) on nucleation in a liquid is not well established. A model was developed for cases where the nucleating phase has a different chemical composition than the initial phase (the coupled-flux model[16]). In this case, the growth of the nuclei leads to a change in chemical composition near the cluster. This interacts with the long-range diffusion field, leading to a coupling of the stochastic processes of interfacial kinetics and diffusion. If the interfacial attachment/detachment rates are faster than the diffusion rate, the kinetics of nucleation can be significantly decreased. The coupled flux model has been explored numerically[17] and has been applied successfully for the precipitation of oxygen in single crystal silicon.[18] For crystal nucleation from a quiescent liquid of a different chemical composition, the coupled fluxes should decrease the nucleation rate. If the liquid is stirred, however, the fluid flow should enhance solute transport and the nucleation rate should increase towards the expected steady-state nucleation rate. To our knowledge a systematic study of the effect of stirring on nucleation of solids from supercooled liquids does not exist.

Here, we report the results of studies of the effect of stirring on nucleation in electromagnetically levitated metallic liquid drops under the microgravity environment of the International Space Station (ISS). Such controlled stirring experiments are not possible under terrestrial conditions because of the natural density and surface tension (Marangoni)[19] driven convection. Microgravity eliminates the first because of the absence of gravity-induced flows. Nearly uniform heating of the sample with a radio-frequency (rf) electromagnetic field in the EML minimizes the second. The EML facility on the ISS, ISS-EML,[20] is capable of controlling the fluid flow by electromagnetic stirring. A brief description of the facility can be found in the Methods section. Three different alloys were chosen to investigate the effect of stirring on nucleation rates in supercooled liquids for different solidification motifs (polymorphic and multi-phase). A $Ti_{39.5}Zr_{39.5}Ni_{21}$ alloy in which primary nucleation is to a quasicrystal of the same composition[21,22] was one choice. It was observed that, although metastable, the quasicrystal phase nucleates first from the supercooled liquid, followed by its decomposition into the stable Laves and solid-solution phases at higher temperatures as the liquid temperature increased due to the release of the heat of fusion. The other two were the bulk metallic glass forming $Cu_{50}Zr_{50}$ and $Zr_{57}Cu_{15.4}Ni_{12.6}Al_{10}Nb_5$ (Vit106) alloys. The primary crystallizing phase from the $Cu_{50}Zr_{50}$ liquid is a cubic B2-phase of $CuZr$,[23] which is consistent with the phase diagram.[24] In contrast, crystalline phases of $Zr_2Ni$, $Zr_2Cu$, $ZrCu$, $Zr_3Al_2$, $Zr_4Al_3$, which also contain various amounts of other elements, crystallize from the glass of Vit106.[25] The results obtained clearly demonstrate that increased stirring increases the nucleation rate for the Vit106, which shifts the crystallization to higher temperatures; little or no effect of stirring was observed for the other two. The results are consistent with the expectations from the coupled-flux model. This is a new observation which is of considerable interest for a fundamental understanding of the nucleation mechanism in partitioning systems. These results could be also vitally important for the future processing of materials in space, which will be necessary for long-term space explorations, and ultimately for manufacturing under extra-terrestrial conditions on Mars and the Moon.

## RESULTS

The experimental details are described in the Methods section of this paper and in refs. [20,26]. Independent control of the heating and positioning of a sample in the ISS-EML is one of the advantages compared to the ground-based EML facilities. Two different electronic circuits feed currents into a pair of identical Cu-coils for sample positioning and heating. The radio frequency (rf)-electromagnetic fields generated by these currents interact with the electrons in the solid/liquid generating eddy currents, which affect fluid flow.[26] Since these rf-fields can be controlled by different combinations of heater and positioner voltages, stirring in liquids can be controlled systematically. Because of the stronger coupling of the heater to the sample, the heater voltage/current is more effective than the positioner in increasing stirring in the liquid.[20,26] The samples were processed under vacuum ($10^{-8}\ mbar$) and in 350 $mbar$ Ar- and He-atmospheres. Figure 1 shows typical time-temperature data for a 6.0 mm diameter Vit106 liquid sphere that was cooled under vacuum from 1440 K, and in a 350 $mbar$ Ar- and He-atmospheres from 1550 K ($T_l = 1123\ K$) with a 2.5 V positioner voltage and the heater off. The sudden rise in temperature at 847 K under vacuum marks the nucleation and growth of the crystal phases from the liquid; the temperature rise is due to the release of heat of fusion during solidification. The onset of the recalescence (nucleation temperature, $T_u$) marks the limit of supercooling (276 K = $0.246 T_l$) for this particular thermal cycle. The cooling rate near the onset temperature for recalescence was 1.3 K/s, which was not sufficient to prevent crystallization. However, the higher cooling rate of 2.2 K/s in the Ar-atmosphere almost suppressed crystallization; only two small kinks near 926 K and 786 K were observed. When processed in He-atmosphere, a much faster cooling rate of 15 K/s was achieved around 850 K, which completely suppressed crystallization, as shown in Fig. 1. This thermal cycle was used for specific heat measurements by modulating the temperature at various hold temperatures. After the last modulation at 1050 K, the heater was turned off for rapid cooling. No crystallization event was observed. That this sample transformed into a glass was evident when it was heated in the next thermal cycle. A sudden rise in temperature due to a glass-crystal transformation was observed between 834 and 922 K, as shown in the inset of Fig. 1. This thermal event is similar to observations made in ground-based ESL studies when a glass was heated rapidly.[27] Unfortunately, subsequent processing of the sample in the He-atmosphere caused contamination, possibly from trace amounts of oxygen in the gas, which prevented glass formation. Our ground-based studies[27] indicated that 500-600 ppm oxygen is enough to prevent bulk glass formation in this alloy. A similar observation of glass formation in a Vit105 ($Zr_{52.5}Cu_{17.9}Ni_{14.6}Al_{10}Ti_5$) alloy under microgravity was reported earlier;[28] this is then the second demonstration that a bulk metallic glass (6.0 mm diameter) can be manufactured in space. Although not unexpected, this result may be of importance for future space-based manufacturing of intricate machine parts using bulk metallic glasses.

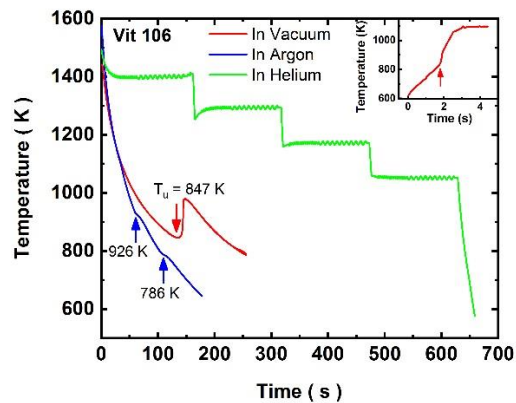

Fig. 1. Typical time-temperature plots during cooling of a Vit106 ($Zr_{57}Cu_{15.4}Ni_{12.6}Al_{10}Nb_5$) liquid under vacuum, Ar- and He-atmospheres. Complete and partial crystallizations happened in vacuum and under Ar-atmospheres. The arrows indicate the crystallization events. Crystallization was completely suppressed in He-atmosphere due to a much faster cooling. The inset shows the recalescence rise in temperature when the glass transformed into crystals during the next heating cycle.

When a 6.5 mm diameter $Cu_{50}Zr_{50}$ liquid sample was cooled in the ISS-EML using a small positioner voltage (3.9V) and heater off under vacuum, the sample undercooled by 252 K ($0.206T_l$, $T_l = 1222$ K), which is less than the value of 322 K observed in ground-based ESL studies of a much smaller sample (2.5 mm diameter).[29] In comparison, a 6.0 mm diameter $Ti_{39.5}Zr_{39.5}Ni_{21}$ liquid in the ISS-EML undercooled by 87 K ($0.082T_l$, $T_l = 1063$ K) for the metastable icosahedral phase under similar conditions because the structure of the icosahedral phase is similar to that of the liquid.[21,30] Note that if the transformation of the icosahedral phase into a mixture of stable Ti-

Zr solid solution and Laves phase can be suppressed, it melts at 1063 K, compared to 1093 K for the phase mixture of the same composition.[31] A slightly higher undercooling of 110 K was possible in the ground-based ESL studies of this liquid.[32] These differences in undercoolability may be due to many factors, such as slower cooling rates (about 3 K/s near $T_u$ compared to about 12 K/s in the ESL) for the larger samples used in space, larger sample volume, larger stirring due to eddy currents induced in the liquid by the EML, and/or small differences in sample purities. Also, heterogeneous nucleation due to small amounts of impurities cannot be completely ruled out.

Since nucleation is a stochastic process, a distribution of nucleation temperatures for a large number of heating and cooling cycles is usually observed.[33] Typically, many hundreds of cycles are required to obtain statistically significant data for this distribution.[29,32,34,35] To obtain a statistically significant result for the effect of stirring, then, many thermal cycles for the same heater and positioner voltages must be performed. However, processing in the ISS-EML is restricted by several considerations. To enhance the life of the facility, each sample is allotted a certain amount of mass loss due to evaporation, part of which deposits on the Cu-coils. In addition to nucleation studies, thermophysical properties (viscosity, surface tension, and specific heat) and electrical properties of the liquids were also measured. This restricted the number of melting cycles from three to five for a particular setting of the heater and positioner voltages. Lower vapor depositions on the Cu-coils in the He-gas atmosphere allowed more thermal cycles (ten to twenty) to be performed. However, once the samples were melted in gas atmospheres, the nucleation temperature started rising and reached a saturation level after a few cycles. Some surface features were also observed in the video images during the cooling of the liquid drops, indicating oxide precipitation. Therefore, the data collected in vacuum are considered to be more representative of homogeneous nucleation than those in the gas atmospheres. For that reason, the data presented here are from vacuum processing only. The experiments were conducted over several nights on the ISS due to restrictions on the allotted times on a given day.

According to the classical nucleation theory,[1] the homogeneous nucleation rate, $I(T)$ at a temperature, $T$, is given by,

$$I(T) = A^* \exp\left(-W^*/k_B T\right), \tag{1}$$

$$\text{where, } W^* = 16\pi\sigma^3/3\Delta G^2, \tag{2}$$

The prefactor, $A^*$ primarily contains kinetic parameters, such as the diffusion coefficient, the atomic jump distance, and the number of possible sites for nucleation; $W^*$ contains the interfacial free energy, $\sigma$, and the driving free energy, $\Delta G$. If the nucleation temperature, $T_u$, is determined from several cycles for a sample of volume $V$, cooled at a rate of $Q$, it follows a Poisson's distribution given by,[33]

$$\omega(t) = \frac{I(T)VT}{Q} \exp\left(-\int_{T_u}^{T_l} \frac{I(T)VT}{Q} dT\right), \tag{3}$$

By fitting the experimentally measured distribution to eqn. (3), the relevant nucleation parameters $A^*, W^*$, and $\sigma$ may be determined. According to the coupled flux model,[16] $A^*$ is the more appropriate parameter for studying the effects of stirring since it contains a combination of the interfacial attachment kinetics and the diffusion coefficient in the liquid. Initially, melting/solidification cycles were performed in vacuum under the most quiescent condition possible (minimum positioner voltage and heater off) and the parameters $A^*, W^*$, and $\sigma$ were determined by fitting to eqn. (3). Thereafter, to investigate the effect of stirring, $\sigma$ was kept constant and $A^*$ was varied to give the best fit to the nucleation data obtained with increased stirring. In all cases the driving free energy was estimated using the Turnbull approximation,[36] $\Delta G = \Delta H_f \Delta T/T_l$, where $\Delta H_f$ is the heat of fusion and $\Delta T$ is the undercooling.

The fluid flow velocities and shear rates corresponding to each positioner and heater voltage settings were simulated following procedures described in refs. [37,38] and in the supplemental section. The fluid flow velocities and shear rates usually show a distribution, which evolves with decreasing temperatures, being largest near the poles and smallest near the equator of the sample due to the distribution of the rf-field along the sample geometry. The parameter used for comparison with the nucleation data are the maximum fluid flow velocities and shear rates at the nucleation temperature.

**Stirring effect on nucleation during processing in vacuum**

Extensive studies of homogeneous nucleation using the ground-based ESL to measure many hundreds of solidification cycles for $Cu_{50}Zr_{50}$[29] and $Ti_{39.5}Zr_{39.5}Ni_{21}$[32] liquids were reported earlier. Similar studies could not be made for the Vit106 alloy since the smaller (~2.5 mm diameter) samples required for the ESL studies formed glasses upon cooling, bypassing crystal nucleation. Unfortunately, as mentioned earlier, such a large number of cycles to obtain statistically significant data for nucleation was not possible with the ISS-EML facility. This prevented a full quantitative analysis of the Poisson's distribution for $T_u$ according to eqn. (3). However, since the cooling rates under different heater and positioner voltages were different, eqn. (3) could still be applied to incorporate those differences. Instead of reproducing the statistical distribution, the focus was to match the average $T_u$ by adjusting $A^*$ and $W^*$. The results from the analysis are listed in Tables 1, 2, and 3 for Vit106 and $Cu_{50}Zr_{50}$ and $Ti_{39.5}Zr_{39.5}Ni_{21}$, respectively. The corresponding estimates for the maximum fluid flow velocities and shear rates are also included.

The values for $A^*$ and $\sigma$ for $Cu_{50}Zr_{50}$ obtained from the ground-based ESL studies[29] were $2.5 \times 10^{31}$ $m^{-3}s^{-1}$ and 0.113 $Jm^{-2}$, which are slightly different from the ISS-based EML studies under the most quiescent conditions (minimum positioner voltage and heater off), as presented in Table 2. The values of $A^*$ and $\sigma$ from the ESL studies for $Ti_{39.5}Zr_{39.5}Ni_{21}$[32] were $2.7 \times 10^{25} m^{-3}s^{-1}$ and 0.057 $Jm^{-2}$, respectively, which are again slightly different from those obtained from the nucleation studies on the ISS under vacuum (Table 3). Such differences may be due to different purities of the samples in the ISS than those used in the ground-based ESL studies. However, since the same starting material and same preparation conditions were used for all alloys prepared for the ISS and ground-based studies, possibly some small amount of contamination in the ISS samples may have occurred during long term storage, transportation (vibrations during rocket acceleration/deceleration rattle the sample in the sample holder), and/or processing on the ISS; fluid flows may also be different under the most quiescent conditions possible in the ground-based ESL and ISS-EML. Considering all possibilities, it is not clear whether the nucleation in these liquids is homogeneous or heterogeneous. Irrespective of that, the ISS data represented in Tables 1, 2, and 3 demonstrate that $A^*$ increased by several orders of magnitude for the Vit106 alloy and remained nearly constant for the other two liquids with increasing stirring. Note that, according to the coupled-flux model,[16] the effect of stirring on nucleation is expected to be observed both for homogeneous and heterogeneous nucleation. The fluid flow calculations were made using the experimental data for the liquid viscosity, which was measured using the ground-based ESL facility, are presented in the supplemental section (Fig. S.1). A typical fluid flow velocity distribution in a Ti$_{39.5}$Zr$_{39.5}$Ni$_{21}$ liquid is also shown in Fig. S.2 for illustration. The errors in those estimates are about ±5%.

To summarize, the average nucleation temperatures, nucleation parameters derived from the ISS-EML data using eqn. (3), corresponding heater and positioner voltages, and the calculated fluid flow velocities and shear rates at the nucleation temperature, $T_u$, for the three alloys are shown in the following three tables. Note that the heater off and heater at 0 V are two different conditions since about 20A current still flows in the coils when the heater is on at 0 V.

Table 1. Vit 106 (Zr$_{57}$Cu$_{15.4}$Ni$_{12.6}$Al$_{10}$Nb$_5$)

| Average nucleation temperature, $T_u$ (K) | Undercooling (K) | Positioner voltage (V) | Heater voltage (V) | Fluid flow velocity at $T_u$ ($ms^{-1}$) | Shear rate at $T_u$ ($s^{-1}$) | Interfacial energy, σ ($Jm^{-2}$) | $A^*(m^{-3}s^{-1})$ |
|---|---|---|---|---|---|---|---|
| 853 | 270 | 2.5 | Off | 1.3x10$^{-7}$ | 3x10$^{-4}$ | 0.12 | 2.5x10$^{35}$ |
| 860 | 263 | 3.4 | Off | 1.9x10$^{-7}$ | 3.9x10$^{-4}$ | 0.12 | 6x10$^{35}$ |
| 872 | 251 | 4.3 | Off | 7.4x10$^{-7}$ | 0.0017 | 0.12 | 7x10$^{38}$ |
| 880 | 243 | 6.0 | Off | 2.2x10$^{-6}$ | 0.0035 | 0.12 | 7x10$^{40}$ |

Table 2. Cu$_{50}$Zr$_{50}$

| Average nucleation temperature, $T_u$ (K) | Undercooling (K) | Positioner voltage (V) | Heater voltage (V) | Fluid flow velocity at $T_u$ ($ms^{-1}$) | Shear rate at $T_u$ ($s^{-1}$) | Interfacial energy, σ ($Jm^{-2}$) | $A^*(m^{-3}s^{-1})$ |
|---|---|---|---|---|---|---|---|
| 970 | 252 | 3.9 | Off | 2.5x10$^{-4}$ | 0.49 | 0.100 | 5x10$^{33}$ |
| 973 | 249 | 4.8 | Off | 3.5x10$^{-4}$ | 0.71 | 0.100 | 2x10$^{34}$ |

Table 3. $Ti_{39.5}Zr_{39.5}Ni_{21}$

| Average nucleation temperature, $T_u$ (K) | Undercooling (K) | Positioner voltage (V) | Heater voltage (V) | Fluid flow velocity at $T_u$ ($ms^{-1}$) | Shear rate at $T_u$ ($s^{-1}$) | Interfacial energy, σ ($Jm^{-2}$) | $A^*(m^{-3}s^{-1})$ |
|---|---|---|---|---|---|---|---|
| 976 | 87 | 3.9 | Off | 3.9x10$^{-4}$ | 0.87 | 0.052 | 2x10$^{23}$ |
| 974 | 89 | 5.7 | Off | 7.8x10$^{-4}$ | 1.7 | 0.052 | 3x10$^{22}$ |
| 975 | 88 | 4.0 | 0.0 | 1.7x10$^{-3}$ | 3.9 | 0.052 | 6x10$^{21}$ |

To demonstrate that the changes in the nucleation parameters are due to stirring, Fig. 2 shows the logarithm of the prefactor, $A^*$ as a function of shear rates for processing in vacuum for the Vit106 liquid. Clearly, $A^*$ increases with the maximum shear rate at the nucleation temperature in the liquid due to electromagnetic stirring. Although $A^*$ correlates well with both the maximum fluid flow velocity and the maximum shear rate at $T_u$ (see Tables 1-3), only the correlation with the shear rate is shown in fig. 2 for clarity. The $Cu_{50}Zr_{50}$ liquid showed a small increase in $A^*$ when the maximum fluid velocity increased from $2.5x10^{-4}$ to $3.5x10^{-4}\ ms^{-1}$ (Table 2). This small increase may not be statistically significant. Therefore, the results are consistent with expectations from the coupled-flux model, since the compositions of the liquid and the primary crystallizing phase are the same. However, more data are needed to make this conclusion statistically significant. Similar consideration applies to the changes in $A^*$ with fluid flows for the $Ti_{39.5}Zr_{39.5}Ni_{21}$ liquid (Table 3). Note that the changes in $A^*$ for the nearly similar values of $T_u$ for this liquid are due to the slower cooling rates for the higher positioner voltages and heater on conditions.

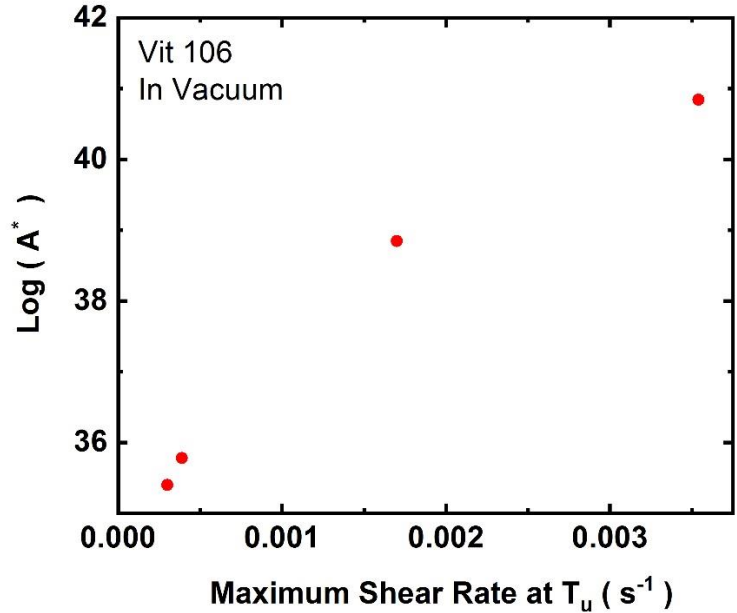

Fig. 2. The prefactor $A^*$ for nucleation as a function of the maximum shear rate at the nucleation temperature ($T_u$) for processing in vacuum for liquid Vit106 ($Zr_{57}Cu_{15.4}Ni_{12.6}Al_{10}Nb_5$). The nucleation temperatures and the relevant parameters are presented in tables 1.

**CONCLUSION**

Two important new results emerge from this study. First, we have confirmed earlier studies,[28] showing that bulk metallic glasses can be synthesized under microgravity. The second and most important observation is that the nucleation rate is significantly enhanced with increasing fluid flow when the composition changes during crystallization (Vit 106), and little or no changes occur for primary crystallization ($Cu_{50}Zr_{50}$ and $Ti_{39.5}Zr_{39.5}Ni_{21}$). For approximately an order of magnitude increase in the fluid flow velocities, the pre-factor, $A^*$, increased by about five orders of magnitude for the Vit106 liquid (Table 1). This occurred even though the fluid flow velocities and the shear rates remained low throughout the range of experiments. For $Cu_{50}Zr_{50}$, $A^*$ increased by about four times for about 40% increase in fluid flow velocity and shear rates. The correlation between $A^*$ and fluid flow is less clear for the $Ti_{39.5}Zr_{39.5}Ni_{21}$ liquid; for three separate fluid flow conditions $A^*$ decreased by a factor of 33 for an increase in fluid flow by about a factor of 4, which could be an artifact due to the lack of statistics. However, the data do clearly suggest that for the $Cu_{50}Zr_{50}$ and $Ti_{39.5}Zr_{39.5}Ni_{21}$ liquids the effect of fluid flow on nucleation is much smaller than for the Vit106 liquid. According to the coupled-flux model,[16] stirring should have a large effect on nucleation when compositional changes occur during solidification, since a concentration gradient is expected to develop in the vicinity of the nucleating cluster, as observed for Vit106. Smaller or no changes should be expected when little or no change in composition occurs during crystallization, as for the polymorphic crystallization for $Ti_{39.5}Zr_{39.5}Ni_{21}$ and $Cu_{50}Zr_{50}$ liquids. The data are then qualitatively consistent with the predictions of the coupled-flux model. A quantitative comparison will require more statistically significant data, obtained from hundreds of nucleation cycles. Future experiments on the ISS are planned for liquids that have lower vapor pressures so that nucleation data from a large number of thermal cycles can be obtained; the higher liquidus temperatures for these liquids will also facilitate higher levels of stirring.

## METHODS

### Sample preparation and nucleation measurements

Nearly spherical samples of $Ti_{39.5}Zr_{39.5}Ni_{21}$ (6.0 mm diameter), $Zr_{57}Cu_{15.4}Ni_{12.6}Al_{10}Nb_5$ (Vit106, 6.0 mm diameter), and $Cu_{50}Zr_{50}$ (6.5 mm diameter) were prepared on Earth by arc melting high purity (> 99.9 at.%) elements in the stoichiometric ratios in a $Ti_{50}Zr_{50}$ gettered high purity (5N) Ar atmosphere. When melted under microgravity, the samples became perfectly spherical. The first two samples were transported to the ISS in metallic cage holders and the last one in a ceramic cup holder. The lower vapor pressure of the $Ti_{39.5}Zr_{39.5}Ni_{21}$, $Zr_{57}Cu_{15.4}Ni_{12.6}Al_{10}Nb_5$ liquids allowed the samples to be studied in the cage holder; the cup was used for the $Cu_{50}Zr_{50}$ to contain evaporation from this higher vapor pressure liquid. The ISS-EML facility[20,26,39] consists of two precisely aligned, identical water-cooled Cu coils. Two separate circuits feed rf voltages to these coils; one for positioning the sample at 150 kHz and the other for heating at 350 kHz. The currents in the two coils for the positioning field were supplied in the opposite directions to produce a quadrupole field. Since the samples were under microgravity, no levitating force was required; a small quadrupolar electromagnetic field provided enough force to keep the sample in position, which caused very limited heating of the sample. Heating currents circulated in the same direction so that a dipole rf electromagnetic field was produced. Unlike most ground-based EML facilities, the heating and positioning can therefore be controlled independently. Processing under high vacuum (~$10^{-8}$ mbar) as well as in high purity Ar- and He-gas atmospheres are possible.

Typically, the samples were heated 300−400 K above $T_l$ and then allowed to cool under various combinations of positioner and heater voltages to induce various levels of stirring/fluid flows. The samples were monitored by two video cameras positioned along the polar and radial directions. The temperatures were monitored by an optical pyrometer (1.45−1.80 $\mu m$) between 573−2373 K. Nucleation and subsequent growth of crystals were marked by a sudden rise in temperature, called recalescence, due to the release of latent heat of fusion. The nucleation temperatures were measured during several melting and solidification cycles under different fluid flow conditions. The measured temperatures were based on emissivity values and electromagnetic coupling coefficients of the samples with the heater and positioner coils, determined from ground-based studies. In actual experiments on the ISS, the sample position shifted, depending on the heater and positioner voltages applied. This changed the measured temperatures, which had to be corrected. This was performed using the measured liquidus temperatures on the ISS-EML and those measured by the DTA technique on ground using a relationship, $1/T_{correct} = 1//T_{measured} + 1/T_{correct\ liquidus} - 1/T_{measured\ liquidus}$, where all temperatures are in K. The experiments on each sample were conducted over several nights. The samples were stored in Helium gas during the intervening periods. In some cases, the nucleation temperatures increased during the initial thermal cycles compared to the previous night's operations. After a few melting cycles it reverted back to the previous values. Therefore, those initial cycles were excluded from the analysis. It was also observed that if the liquid was disturbed during cooling by pulsing for the viscosity measurements, or the positioner/heater voltages were changed during cooling, the nucleation temperature changed. For those reasons, the nucleation data from those cycles were also excluded. The reason for such changes is not clear at the moment. Maximum undercooling for the $Ti_{39.5}Zr_{39.5}Ni_{21}$ liquid was achieved after many melting cycles under vacuum over several nights. Possibly, the liquid purified during such repeated melting. The data presented here are from these later melting cycles when consistent maximum undercoolings could be achieved.

**Model calculations for fluid flow**

Fluid flow velocities and shear rates are the two parameters that give a direct measure of the stirring. Since they cannot be measured directly in the present experimental arrangement, model calculations were necessary. The voltages and currents applied to the positioner and heater coils, their frequencies, the position and diameter of the sample with respect to the coils, sample electrical resistivity and coupling coefficient to the rf fields, the liquid viscosity and density as a function of temperature are the input parameters in this model.[37,38] The particular details of the coil geometry and the rf fields can be found elesewhere.[39] The temperature dependent viscosities of the liquids were measured by a ground-based electrostatic levitation facility[40,41] at Washington University in St. Louis and were reported earlier.[42,43] A commercial software, ANSYS, was used to solve the Navier-Stokes equation under the conditions of the experiment. The model used was verified from an earlier experiment on the ISS where the flow rates were measured from video images of particles flowing on the surface of an immiscible CuCo liquid;[44] the model calculations agreed with the experimental results within 7%.

## DATA AVAILABILITY

The experimental data are available from the corresponding author on reasonable request.

# REFERENCES

1. Kelton K. F. and Greer A. L. *Nucleation in Condensed Matter* (Elsevier, Boston, 2011).
2. Glicksman M. E. *Principles of Solidification* (Springer, Berlin, 2011).
3. Herlach D. M., Galenko P., and Holland-Moritz D., *Metastable solids from undercooled melts,* Pergamon Materials Series, edited by R.W. Cahn, Elsevier (2007).
4. Rhim W.-K., Collender M., Hyson M. T., Simms W. T., Elleman D. D. "Development of an electrostatic positioner for space material processing." *Rev. Sci. Instrum*. **56**, 307 (1985).
5. Herlach D. M., Cochrane R. F., Egry I., Fecht H.-J., Greer A. L. "Containerless processing in the study of metallic melts and their solidification". *Int. Mat. Rev*. **38**, 273 (1993).
6. Weber J. K. R., Hampton D. S., Merkley D. R., Rey C. A., Zatarski M. M., Nordine P. C. "Aeroacoustic levitation: A method for containerless liquid phase processing at high temperatures". *Rev. Sci. Instrum*. **65**, 456 (1994).
7. Trinh E. H. "Compact acoustic levitation device for studies in fluid dynamics and materials science in the laboratory and microgravity". *Rev. Sci. Instrum*. **56**, 2059 (1985).
8. Wedekind J., Hyvärinen A.-P., Brus D., and Reguera D. "Unraveling the pressure effect on nucleation". *Phys. Rev. Lett*. **101**, 125703 (2008).
9. Karpov I. V., Mitra M., Kau D., Spadini G., Kryukov Y. A., and Karpov V. G. "Evidence of field induced nucleation in shape memory". *J. Appl. Phys.* **92**, 173501 (2008).
10. Hou H. and Chang H.-C. "ac field enhanced protein crystallization". *Appl. Phys. Lett*. **92**, 223902 (2008).
11. Gavira J. A. and Garcia-Ruiz J. M. "Effects of a magnetic field on Lysozyme crystal nucleation and growth in a diffusive environment". *Cryst. Growth & Design*. **9(6)**, 2610 (2009).
12. Schenk T., Holland-Moritz D., and Herlach D. M. "Observation of magnetically induced crystallization of undercooled Co-Pd alloys". *Europhys. Lett*. **50**, 402 (2000).
13. Holland-Moritz D. and Spaepen F. "The magnetic contribution to the driving force for crystal nucleation in undercooled Co-Pd melts". *Phil. Mag*. **84**, 957 (2004).
14. Galenko P. K., Reuther K., Kazak O. V., Alexandrov D. V., and Rettenmayr M. "Effect of convective transport on dendritic crystal growth from pure and alloy melts". *Appl. Phys. Lett*. **111**, 031602 (2017).
15. Galenko P. K., Wonneberger R., Koch S., Ankudinov V., Kharanzhevskiy E. H., Rettenmayr M. "Bell-shaped "dendrite velocity-undercooling" relationship with an abrupt drop of solidification kinetics in glass forming Cu-Zr(-Ni) melts". *J. Crystal Growth* **532**, 125411 (2020).
16. Kelton K. F. "Time-dependent nucleation in partitioning transformations". *Acta Mater*. **48**, 1967 (2000).
17. Diao J., Salazar R., Kelton K. F., Gelb L. D. "Impact of diffusion on concentration profiles around near-critical nuclei and implications for theories of nucleation and growth". *Acta Mater*. **56**, 2585 (2008).

18. Wei P. F., Kelton K. F., and Falster R. "Coupled-flux nucleation modeling of oxygen precipitation in silicon". *J. Appl. Phys.* **88**, 5062 (2000).
19. Getting A. V. *Rayleigh-Benard convection: structures and dynamics* (World Scientific, Singapore, 1988).
20. Seidel, A., Soellner, W. and Stenzel, C. "EML—An electromagnetic levitator for the International Space Station". *J. Phys. Conf. Ser.*, **327**, 012057 (2011).
21. Kelton K. F., Lee G. W., Gangopadhyay A. K., Hyers R. W., Rathz T. J., Rogers J. R., Robinson M. B., and Robinson D. S. "First X-Ray Scattering Studies on Electrostatically Levitated Metallic Liquids: Demonstrated Influence of Local Icosahedral Order on the Nucleation Barrier". *Phys. Rev. Lett.* **90**, 195504 (2003).
22. Lee G. W., Croat T. K., Gangopadhyay A. K., and Kelton K. F. "Icosahedral-phase formation in as-cast Ti-Zr-Ni alloys". *Phil. Mag. Lett.* **82**, 199 (2002).
23. Gegner J., Shuleshova O., Kobold R., Holland-Moritz D., Yang F., Hornfeck W., Bedarnick J., Herlach D. M. "In situ observation of the phase selection from the undercooled melt in Cu-Zr". *J. Alloy. Comp.* **576**, 232 (2013).
24. Okamoto H. "Cu-Zr phase diagram". *J. Phase Equilibria and Diffusion* **29**, 204 (2008).
25. Shadowspeaker L., Shah M., Busch R. "On the crystalline equilibrium phases of the $Zr_{57}Cu_{15.4}Ni_{12.6}Al_{10}Nb_5$ bulk metallic glass forming alloy". *Scripta Metal.* **50**, 1035 (2004).
26. Lohoefer G. "High-resolution inductive measurement of electrical resistivity and density of electromagnetically levitated liquid metal droplets". *Rev. Sci. Instrum.* **89**, 124709 (2018).
27. Bendert J. C., Blodgett M. E., Gangopadhyay A. K., and Kelton K. F. "Measurements of volume, thermal expansion, and specific heat in $Zr_{57}Cu_{15.4}Ni_{12.6}Al_{10}Nb_5$ and $Zr_{58.5}Cu_{15.6}Ni_{12.8}Al_{10.3}Nb_{2.8}$ liquids and glasses". *Appl. Phys. Lett.* **102**, 211913 (2013).
28. Mohr M., Wunderlich R. K., Hofmann D. C. and Fecht H.-J. "Thermophysical properties of liquid $Zr_{52.5}Cu_{17.9}Ni_{14.6}Al_{10}Ti_5$—prospects for bulk metallic glass manufacturing in space". *npj Microgravity* **5**, 24 (2019).
29. Sellers M. E., "Studies of maximum supercooling and stirring in levitated liquid metal alloys". Ph. D. Thesis (Washington University in St. Louis, May 2020).
30. Schenk T., Holland-Moritz D., Simonet V., Bellisent R., and Herlach D. M. "Icosahedral short range order in deeply undercooled metallic melts". *Phys. Rev. Lett.* **89**, 075507 (2002).
31. Lee G. W., Gangopadhyay A. K., and Kelton K. F. "Phase diagram studies of Ti-Zr-Ni alloys containing less than 40 at.% Ni and estimated critical cooling rate for icosahedral quasicrystal formation from the liquid." *Acta Mater.* **59**, 4964 (2011).
32. Sellers M. E., Van Hoesen D. C., Gangopadhyay A. K., and Kelton K. F. "Maximum supercooling studies in $Ti_{39.5}Zr_{39.5}Ni_{21}$, $Ti_{40}Zr_{30}Ni_{30}$, $Zr_{80}Pt_{20}$ - Connecting Liquid Structure and the Nucleation Barrier". *J. Chem. Phys.* **150**, 204510 (2019).

33. Skripov V. P. "Homogeneous Nucleation in Melts and Amorphous Films," *Current Topics in Materials Science,* Crystal Growth and Materials, Vol. 2, ed. E. Kaldis and H. J. Scheel. North-Holland, New York, 1977, pp.328-378.
34. Morton C. W., Hofmeister W. H., Bayuzick R. J., Rulison A. J., and Watkins J. L. "The kinetics of solid nucleation in Zirconium." *Acta Mater.* **46**, 6033 (1998).
35. Klein S., Holland-Moritz D., and Herlach D. M. "Crystal nucleation in undercooled liquid Zirconium." *Phys. Rev. B* **80**, 212202 (2009).
36. D. Turnbull, "Formation of crystal nuclei in liquid metals". *J. Appl. Phys*. **21**, 1022 (1950).
37. Hyers R. W., Matson D. M., Kelton K. F., and Rogers J. R. "Convection in containerless processing". *Ann. N. Y. Acad. Sci.* **1027**, 474 (2004).
38. Bracker G. P., Baker E. B., Nawer J., Sellers M. E. Gangopadhyay A. K., Kelton K. F. Xiao X., Lee J., Reinartz M., Burggraff S., Herlach D. M., Rettenmayr M., Matson D., Hyers R. W. "The effect of flow regimes on the surface oscillations during electromagnetic levitation experiments". *High Press. High Temp.* **49**, 49 (2020).
39. Lohöfer, G. & Piller, J. "The new ISS Electromagnetic Levitation Facility—'MSLEML' in *40th AIAA Aerospace Sciences Meeting & Exhibit*, 2002.
40. Gangopadhyay A. K., Lee G. W., Kelton K. F., Rogers J. R., Goldman A. I., Robinson D. S., Rathz T. J., and Hyers R. W. "Beamline electrostatic levitator for in-situ high energy x-ray diffraction studies of levitated solids and liquids". *Rev. Sci. Instr*. **76**, 073901 (2005).
41. Mauro N.A. and Kelton K. F. "A highly modular beamline electrostatic levitation facility, optimized for in situ high-energy x-ray scattering studies of equilibrium and supercooled liquids". *Rev. Sc. Instrum.* **82,** 035114 (2011).
42. Bradshaw R. C., Arsenault A. D., Hyers R. W., Rogers J. R., Rathz T. J., Lee G. W., Gangopadhyay A. K., and Kelton K. F. "Nonlinearities in the undercooled properties of $Ti_{39.5}Zr_{39.5}Ni_{21}$". *Phil. Mag.* **86 (3-5)**, 341 (2006).
43. Blodgett M. E., Egami T., Nussinov Z., and Kelton K. F. "Proposal for universality in the viscosity of metallic liquids". *Sci. Rep.* **5**, 13837 (2015).
44. Lee J., Matson D. M., Binder S., Kolbe M., Herlach D. M., and Hyers R. W. "Magnetohydrodynamic modeling and experimental validation of convection inside electromagnetically levitated Co-Cu droplets". *Metall. Mater. Trans. B*. **45**, 1018 (2014).

## ACKNOWLEDGEMENTS

Work at Washington University was supported by NASA under grant Nos. NNX10AU19G and NNX16AB52G. Work at University of Massachusetts was supported by NASA under grant no NNX16AB40G and 80NSSC21K010. Work at the Otto-Schott-Institut für Materialforschung at Friedrich Schiller University, Jena, was done under the project MULTIPHAS supported by the German Space Center Space Management under contract number 50WM1941. The authors acknowledge the access to the ISS-EML, which is a joint undertaking of the European Space Agency (ESA) and the DLR space administration. The work on the $Ti_{39.5}Zr_{39.5}Ni_{21}$ sample was conducted within the framework of the ESA research project ICOPROSOL (AO-2009-959). The Vit106 and $Cu_{50}Zr_{50}$ alloys also come within the framework of ESA-sponsored projects. We are particularly indebted to the members of the Microgravity Users Support Center (MUSC) at the DLR, Köln, for their generous support in the planning and execution of the experiments.

## AUTHOR CONTRIBUTIONS

K.F.K. conceived the idea. A.K.G., D.H.M., S.K., P.K.G. and K.F.K. planned the experiments. A.K.G. and S.K. prepared the samples. A.K.G., D.H.M., D.C.V, and K.F.K. performed the experiments in cooperation with the members of the Microgravity Users Support Center (MUSC) at the DLR, Köln. A.K.G., M.E.S., and D.H.M. analyzed the data. G.P.B., A.K.P., and R.W.H. performed the fluid flow calculations. A.K.G. and K.F.K wrote the manuscript with inputs from all coauthors.

## COMPETING INTERESTS

The authors have no competing interests.

## MATERIALS & CORRESPONDENCE

All correspondence should be addressed to A.K.G. at <anup@wuphys.wustl.edu>

**SUPPLEMENTAL**

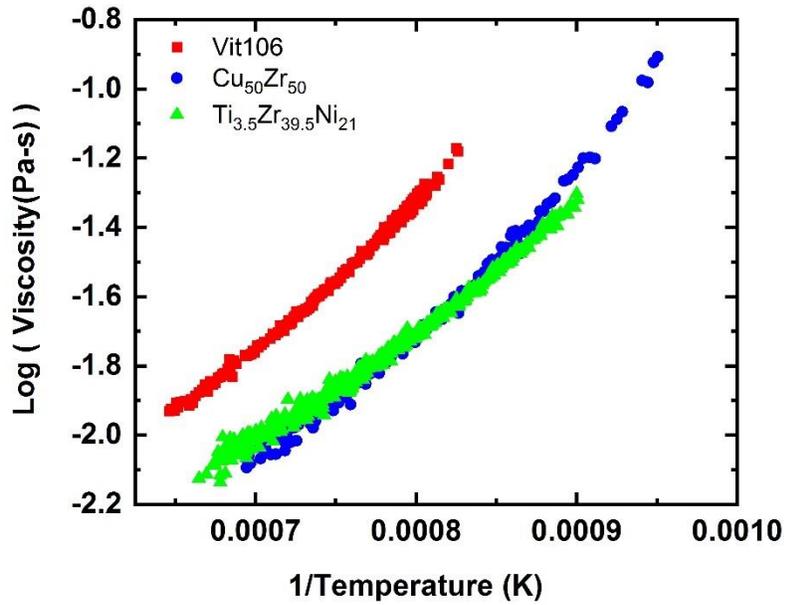

Fig. S.1. The viscosities of three different liquids as a function of temperature measured by the ESL technique.

The fluid flows and shear rates in the liquids were estimated following procedures described in refs. [37,38]. Figure S.2 shows the results for fluid flows for one such calculation for the $Ti_{39.5}Zr_{39.5}Ni_{21}$ liquid at the nucleation temperature of 990 K, when it was cooled in vacuum with 5.7 V positioner and heater off condition.

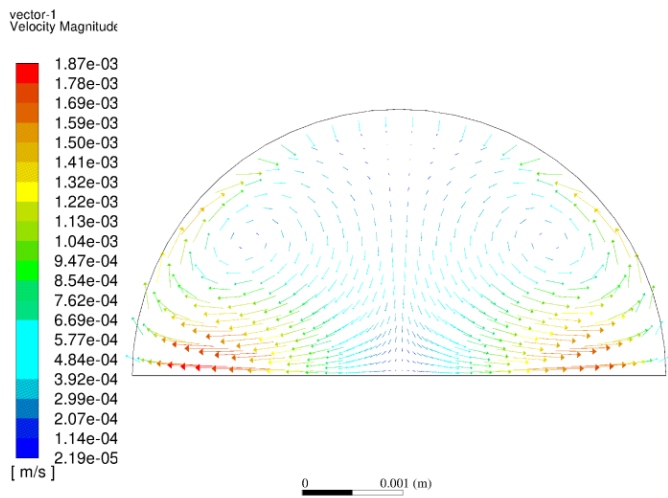

Fig. S.2. The fluid flow velocity distribution in a 6.0 mm diameter $Ti_{39.5}Zr_{39.5}Ni_{21}$ liquid at 990 K when the sample was cooled with a 5.7 V positioner and heater off condition. As is apparent from the color code, the flow velocities are maximum at the equator and minimum at the poles.